# The Elemental Analysis of Elymais Period Coins by PIXE Technique and Their Statistical Analysis by SPSS

Behzad Hosseini Sarbishe

PhD. Candidate in Archaeology,Tarbiat Modarres University

Iran, Tehran

Ali Asghar Salahshoor*

PhD. Candidate in Archaeology, Tehran university

Iran, Tehran

E-mail: aa.salahshour@ut.ac.ir (Corresponded Author)

Hamzeh Ghobadi Zadeh

PhD. Candidate in Archaeology, Tehran University

Iran, Tehran

# The Elemental Analysis of Elymais Period Coins by PIXE Technique and Their Statistical Analysis by Spss Software

**Abstract:**

The Elymaean have existed as a semi-autonomous satrap and sometimes autonomous, coinciding with the Seleucids and the Parthians in the southwest and west of Iran. This government has played an active presence in political and economic events of the region from the mid-second century BC to the advent of the Sasanid Empire. In this paper is tried using of archaeometry matter to Study historical coins of Elymaean. This method is a valuable criterion for recognition of size and weight of coins. Elemental studies of coins give us valuable information in connection with particle size, combinations percentage of the coins, and applied metallurgy in the manufacture of these coins. These studies provide the groundwork for the understanding of the economic and political structure and commercial communications of the society. The difference in the level of element concentration may contain information concerning the mine of these metals. In this paper have been studied total coins of bronze and silver belonging to the Elymaean using PIXE Technique to be can achieve the beneficial results about the identification of concentration of present elements, the number of mines and the number of mints through elemental analysis and definition of the chemical composition of metals. To achieve the above-mentioned goals, we analyzed 35 coins of Elymais. These coins involve three periods of Elymaean ( from 85 BC to 224 AD). The observations results showed that the percentage of main elements of coins such as silver and copper in the three periods of Elymaean were different together and coins of the first period have had more purity. Furthermore, the analyzed coins are minted in six mints and their primal raw materials are from three different mines. Lastly, statistical analysis in this study is performed with SPSS software.

**Keywords**: Elymais Coins, Ore, PIXE technique, SPSS, Elemental Analysis

**Introduction:**

The Elymaean are survivors of Elamites that they governed in the mid-second century BC in the southwest of Iran. The main base of this government was located in the current province of Khuzestan. According to the ancient texts can be noted that they lived in the northeast and southern Khuzestan. The founder of Elymais state was Kamnaskires I who based on the available resources is governed periods between 160 and 140 BC. Elymais state until 224 AD (i.e., contemporary with the beginning of the Sasanid Empire) has had active participation in the regional events (Alizadeh, 1985: 181-183). According to the written sources, the archaeological evidence and the late field research can be limited Elymaean from the north to parts of Khuzestan and Chaharmahal and Bakhtiari provinces, from the east to Kohgiluyeh and Boyer-Ahmad province, from the south to the Persian Gulf and from the west to the plains of Khuzestan province (Moradi, 2012; Hojabri Nobari et al., 2013: 60, 61).

Among the archaeological resources, the coins have particular importance because the coins have high resistance against damage, as well as the small size of this archaeology data which can be shown many information from the cultural, economic, and political conditions of every period. With the typology of coins that is a common technique in the recognition of this cultural data, we can acquire information about the relative chronology, art cognition, etc. Laboratory methods such as PIXE can provide additional information for our knowledge of the target society.

With the advent of the modern archeology, to the study of ancient data is used experts from other sciences such as Archaeometry. One of the non-destructive analytical techniques in the archaeometry is PIXE. This method because of fast and sensitivity is used to study of coins. Today, the PIXE technique, concerning its high accuracy and non-destructive nature is considered in the studies of archaeometry (Tripathy et al., 2010). Benefits of the laboratory methods such as PIXE can be provided additional information for the understanding of the target society. Elemental analysis of metals provides valuable information about the composition of the alloys and the ways of the recognition of the source of metals (Guerra, 2008). In the numismatic, it is important to recognize original compositions of alloy to the cognition of the main alloy of the coins. The relative proportion of the content of the main elements provides valuable information about the changes in the monetary system, economic and political conditions and technology of coinage (Beck et al., 2004). In this paper, we try to study 35 coins of the various kings of Elymais by PIXE technique. Using this way, we can determine the number of mines and the used mint to produce these coins and the analysis of the economic and, political conditions. Based on the coin evidence can be mentioned that this rule can be studied in three time periods. The first period, the family of Kamnaskires, the second period, the government that knows it from the Parthian dynasty (25- 150 AD), and the third period is known as the last kings' state Elymais (150-221 AD).

**Research history:**

The first scholar who studied the pre-Islamic coins of Iran was the American chemist, Caley (1950) who noted the reign of Orodes, king of Parthian. Hughes (Hughes & Hall 1979) not only has worked on Sasanian silver metals but has also compared them with Roman silver metals. Among researchers that using XRF and PIXE techniques have studied available elements in coins or metals of different historical periods can be mentioned, Khademi Nadooshan (2006) and Hajivaliei (2009). Hajivaliei by PIXE technique surveyed the number of the silver coins of Khosrow II, the Sasanid king (Hajivaliei 2008). Among those who have analyzed the copper coins can be mentioned Vijayan, Hajivaliei and Kallithrakas- Kontos. Vijayan studied the copper coins of Kushan government (Vijayan, 2005: 121-125), Hajivaliei the ancient Indian coins (Hajivaliei et al., 1999) and, Kallithrakas- Kontos (1996) analyzed the copper coins of ancient Greece, but the coins of Elymais have not been analyzed, So this study is the first lab activity on the coins of this period. Coins of this period, are mostly copper, silver coins percentage was much lower than the copper and the gold coins have not been achieved.

**Elymaean coins and the mints:**

To a better understanding of the different periods, coins regard to the kings, the sequence of that period is necessary. The newest division of Elymaean has done by Pakzdaidan. He has investigated the Elymais period based on the coins in three parts (Pakzadian, 2007). Coins provide us the most important information for understanding the sequence of Elymais kings. Based on the designs of the coins, alloy, weight and the concentration of different elements can be acquired useful information from this source about economic, religious, and political. Elymais coins were minted from silver, bronze and, copper. The special monetary unit has not been presented for this period, but for other periods such as Parthian, was drachms (tetradrachms and one drachm) which was divided into smaller units such as the smaller sub-units “Aboul” (three Aboul, two Aboul, one Aboul) and also “Kalkon” (like eight Kalkon, four Kalkon, two Kalkon and, one Kalkon), in this way, each one drachm was equivalent six Aboul and each one Aboul was equivalent eight Kalkon. The normal weight of tetradrachms coins during the Parthian period was 14-16 grams, Also one drachm was 3/5-5 grams and Aboul was one-sixth drachms. Bronze coins in through of Parthian government period and across the country were minted in different weights from 7.0 g to 14 g (Sarfaraz & Avarzamani, 2006). In this study is considered silver coins of Elymais as the drachms and bronze coins as the Kalkon.

Elymaean coins from the King, Kamnaskires I (147 BC) to King, Kamnaskires III (ca. before years of 85BC) are derived from the tradition of the Seleucids and from the king, Kamnaskires IV (82- 72 BC) until the end of this government had resembled the Parthian coins. On the obverse of coins are depicted designs and political symbols like king image and related symbols with Elymais kings (in many cases, these symbols are not interpreted correctly). On the reverse of coins have been used various images such as the previous king or crown prince of Shah, parallel and intersecting Scattered lines, birds, etc. At the first, Greek script, and then Aramaic script (including the name of the king, sometimes king name and his epithets) are used. Also mints information and date of mintage are carved on the reverse of some of these coins (Pakzadian, 2007).

Before the Kamnaskires III, Elymais coins are similar to the Seleucids coins with Greek script, but from the Kamnaskires III Government changed pictures of Elymais coins, and tendency to the Parthian iconography is seen on the coins of Elymais that same style continued until the end of the Elymaean government, but script until kings of the second period of Elymais has been Greek which gradually changed to Aramaic script. During the reign of Farhad (Phraates) IV and Orodes IV, contemporary with Kamnaskires XI and XIII, Parthian mints in Susa reduced the value of the drachms, This action affected the coins of Kamnaskires dynasty and this coins from tetradrachms merely had decreased to the bronze sign (Ibid). Kamnaskires- ruler, the founder of the third period of Elymaean government, almost in the second decade of the second century AD, perfectly joined together Susa and Elymais once again. The coinage accuracy and used images on the coins of the second and third Elymais decreased toward the first period of Elymais. Coins of the Elymais kings have been scattered in a wide range of Khuzestan, Lorestan, Chaharmahal and Bakhtiari, Kohgiluyeh and Boyer-Ahmad, Bushehr provinces, Fars province to the east of Saudi Arabia that this subject reflects the important role of Elymaean in the regional and trans-regional trade. Regional developments, particularly the conflicts between the Parthian and the Roman were effective for the economic situation of Elymaean. As far as the sources have pointed out, Elymaean capitals have been Ayapir (Izeh), Susa (Seleucia-by-the-Eulaios), Seleucia-by-the-Hedyphon. The reason for this change is not yet known, but Susa as the key city has been important in all times of the Elymais government. When Susa was dominated by Parthian and Seleucids, their capital was moved to another city (Yarshater, 1989; Pakzadian, 2007). By investigation of written signs on the coins of Elymais, at least is detected name and sign of seven cities of Elymais.

In the early government of Elymais, Ayapir and Susa mints have been active. After decades of sources silence, Elymaean in another place, probably in the first half of the first century BC have minted coin in the Seleucia-by-the-Hedyphon again and the capital was moved there. After the Kamnaskires III, a group of coins has been affected by Parthian coins and, apparently, former Mint of Ayapir has been closed or dominated by the Parthians.

The Elymais kings coins have only been minted Susa after the year 75 AD, but before, these coins were minted in Seleucia-by-the-Hedyphon and imply that the capital was probably moved from Seleucia to Susa in 75 AD. The Elymaean territory during the government of several hundred years has constantly been changed from its size and extent of view. Mints signs on the coins of Elymais are various, but their position is not often specified. By the study of the written signs on the Elymaean coins, at least is identified the name and signs of seven cities. Until before the Kamnaskires IV and Anzaze queen, mints of Ayapir and Susa were main Elymais mints. From this date onwards Susa and Seleucia-by-the-Hedyphon have been the main mints (Vanden Berghe and Shippman, 1985; Pakzadian, 2007). Unfortunately, on the first coins of Elymais Kings have not been found mintage date and on the coins of contemporary with the Parthian coins are rarely seen dating on the coins. Method of writing in these coins like the Seleucid and Parthian coins are carved with the Latin alphabet and its reading is from left to right. For example, ΕΛΣ date based on the table 1 equal to 235 Seleucid or 78 Parthian BC (Pakzadian, 2007)

| | | | | | | | | | | | | | | |
|---|---|---|---|---|---|---|---|---|---|---|---|---|---|---|
| Α | Β | Γ | Δ | Ε | Σ | Ζ | Η | ⊙ | Ι | Κ | Λ | Μ | Ν | Ξ |
| 1 | 2 | 3 | 4 | 5 | 6 | 7 | 8 | 9 | 10 | 20 | 30 | 40 | 50 | 60 |
| Ο | ΙΙ | Q | Ρ | ∑ | Τ | Υ | Φ | Χ | Ψ | Ω | | | | |
| 70 | 80 | 90 | 100 | 200 | 300 | 400 | 500 | 600 | 700 | 800 | | | | |

Table. 1: The words and numbers in the coins dating of Elymais (Pakzadian, 2007)

**Samples preparation**:

The Analysis with the PIXE technique is one of the most accurate analyses for the finding of the small elements. This study used the PIXE technique for the elemental analysis of Elymais coins. By studying the composition and metal content of the coins can be explained and analyzed historical events, place of mint, and identification of their mines. As well as with the achievement of the detectable relations can be recognized the fake and imitation coins or suspicious coins. The studied coins of this research are selected from the personal collection of Mohammad Saffar. The selected collection includes 2 silver coins and 33 copper coins which by authors are placed in acetone for 24 hours and then have been cleared with distilled water. The collection of coins, after dried in the open air, were transferred to the “Van de Graaff” lab of Atomic Energy Organization of Iran and were analyzed by PIXE technique.

We performed the spectrum analysis using Gupix software that offers us a parametric method for quantitative analysis and routinely is used to analyze the whole of PIXE Spectrums. To analyze, we should identify the target Matrix. The matrix means that the highest percentage is related to which element. In this paper, the largest element is silver. In the final, three to five percent error, caused by a fundamental parameter of calibration and lack of cleanliness of the coin is expected (Hajvalaiei, 2009).

The studied coin in this research is tested by PIXE of Atomic Energy Organization of Iran. Then, in order to determine the technology of silver extraction from copper and zinc mines, and the differences in its extraction during the Seleucid and Parthian kingdoms, the results of the spectrometry were processed with SPSS statistical method.

**Discussion:**

In this paper, 35 Elymais coins were analyzed by PIXE technique (Table. 2). Because of the long period of Elymaean government, the various kings, lack of the precise date, and sequence of king's government date, We can not examine all of the coins, but it is tried to analyze samples of different coins of this government to achieve the desired results. The number of 2 silver coins and 33 copper coins were analyzed. After the analysis, it was determined that from these 33 coins, seven numbers are Bilon and 26 coins are copper. The oldest analyzed coin in this study is related to Kamnaskires IV and Queen Anzaze (82 / 81- 72/71 BC) and also the newest analyzed coin is related to the seventh unknown king of the Elymaean government (ca. the late first century or early the second century AD). Four analyzed coins of 35 coins weighed as much as tetradrachms, but they were not made of silver. Two coins were copper and two coins were Bilon. These coins are shown in Figure 1, and Tables 3, 4. The approximate weight of these coins was between 12 and 16 grams. The weight of some silver coins and some Bilon coins were less than 2 grams. The weights of other silver and Bilon coins were 2- 4 grams. The weight of the analyzed Elymais coins is shown in Figure 2.

<table>
<tr><th>No</th><th>Reverse</th><th>Obverse</th><th colspan="6">Explanations of coins</th></tr>
<tr><td rowspan="2">1</td><td rowspan="2">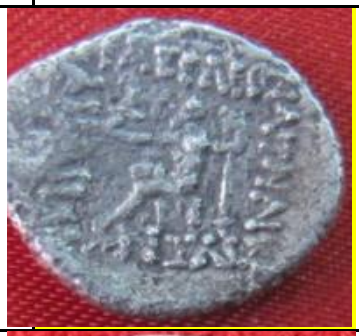</td><td rowspan="2">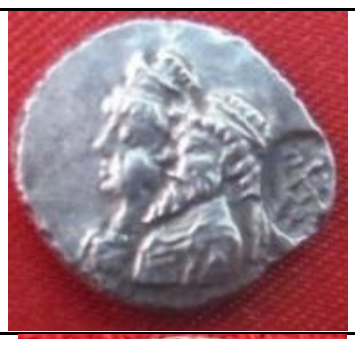</td><td>reference</td><td>mint</td><td>Weight (gr)</td><td>s</td><td>material</td><td rowspan="2">Coin of Kamnaskires IV and Anzaze queen A</td></tr>
<tr><td>Pakzadian 2007</td><td>Kangavar</td><td>1/567</td><td>1.7</td><td>silver</td></tr>
<tr><td rowspan="2">2</td><td rowspan="2">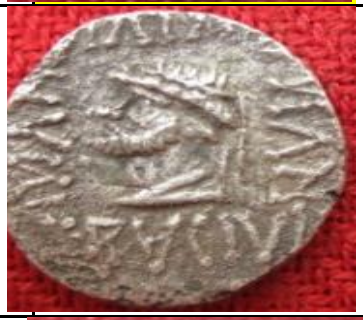</td><td rowspan="2">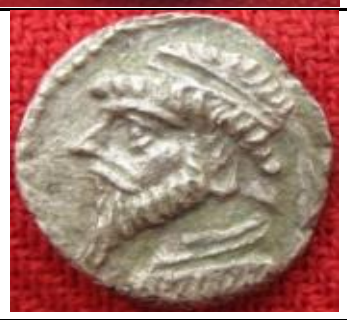</td><td>reference</td><td>mint</td><td>Weight (gr)</td><td>Diameter (cm)</td><td>material</td><td rowspan="2">Coin of Kamnaskires VIII I</td></tr>
<tr><td>Ibid</td><td>?</td><td>3/468</td><td>2</td><td>silver</td></tr>
<tr><td rowspan="2">3</td><td rowspan="2">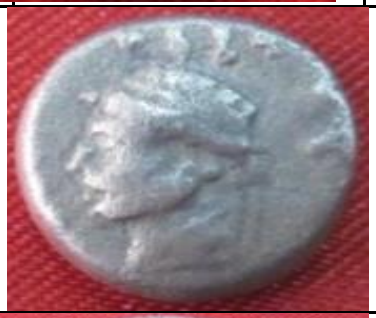</td><td rowspan="2">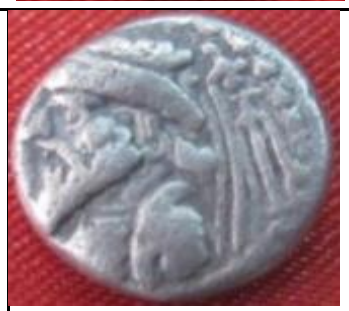</td><td>reference</td><td>mint</td><td>Weight (gr)</td><td>Diameter (cm)</td><td>material</td><td rowspan="2">Coin of Kamnaskires XI O</td></tr>
<tr><td>Ibid</td><td>?</td><td>1/835</td><td>1.8</td><td>Bilon</td></tr>
<tr><td rowspan="2">4</td><td rowspan="2">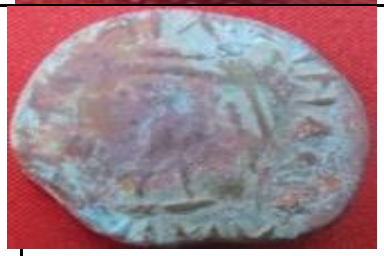</td><td rowspan="2">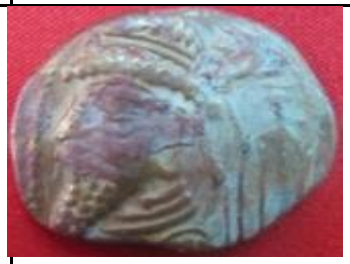</td><td>reference</td><td>mint</td><td>Weight (gr)</td><td>Diameter (cm)</td><td>material</td><td rowspan="2">Coin of Kanaskires XII V</td></tr>
<tr><td>Ibid</td><td>?</td><td>12/839</td><td>3.2</td><td>Bilon</td></tr>
<tr><td rowspan="2">5</td><td rowspan="2">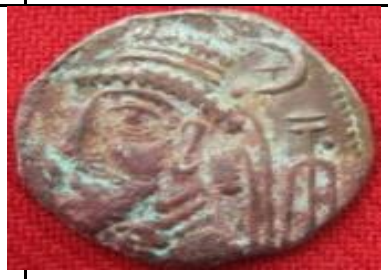</td><td rowspan="2">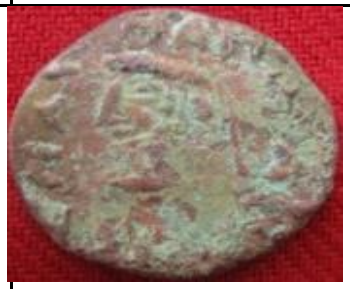</td><td>refernce</td><td>mint</td><td>Weight (gr)</td><td>Weight (cm)</td><td>material</td><td rowspan="2">Coin of Kanaskires XIII T</td></tr>
<tr><td>Ibid</td><td>?</td><td>15/518</td><td>3.3</td><td>Bilon</td></tr>
</table>

| No. | Obverse | Reverse | reference | mint | Weight (gr) | Diameter (cm) | material | Coin |
|---|---|---|---|---|---|---|---|---|
| 6 | 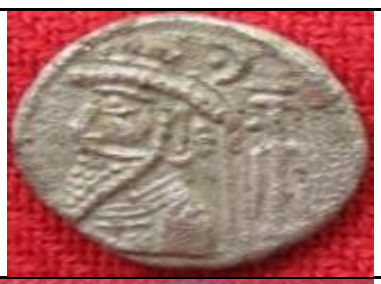 | 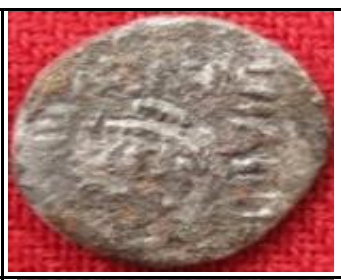 | Ibid | ? | 3/324 | 1.9 | Bilon | Coin of Kamnaskires XII G |
| 7 | 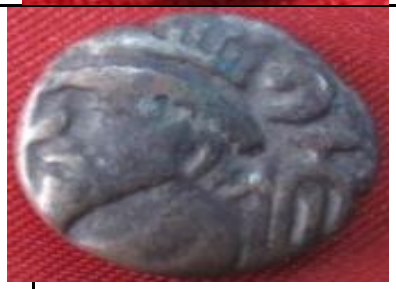 | 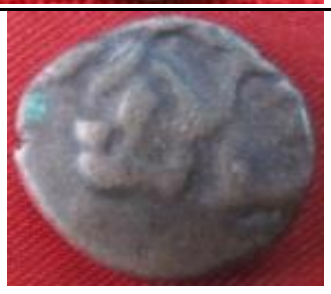 | Ibid | ? | 3/52 | 1.8 | copper | Coin of other kings of Kamnaskires dynasty F |
| 8 | 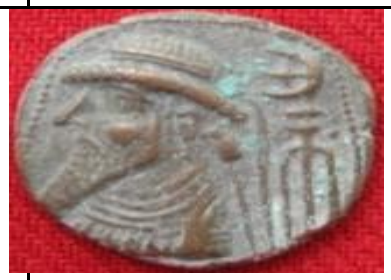 | 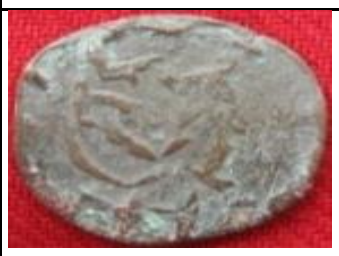 | Ibid | ? | 14/691 | 3.2 | Bilon | Coin of other kings of Kamnaskires dynasty S |
| 9 | 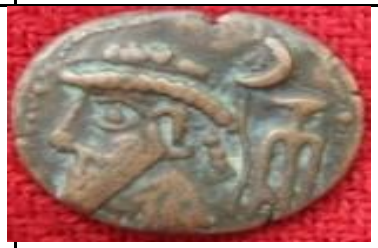 | 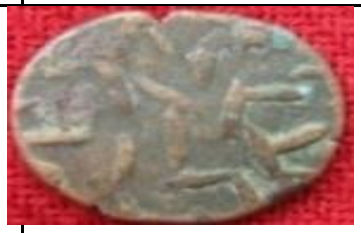 | Ibid | ? | 3/392 | 2.1 | copper | Coin of other kings of Kamnaskires dynasty R |
| 10 | 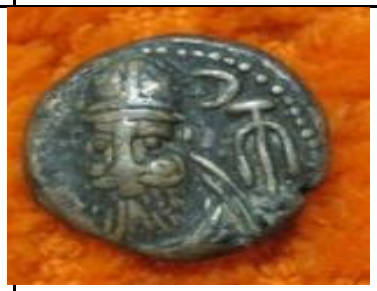 | 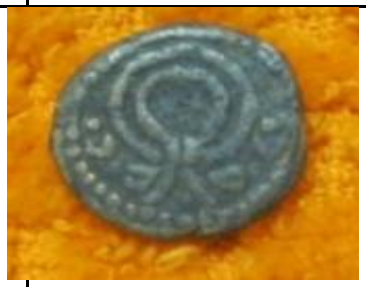 | Ibid | ? | 2/936 | 1.9 | copper | Coin of Phraates I C3 |
| 11 | 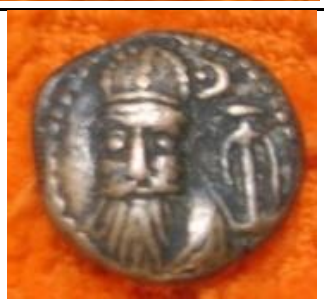 | 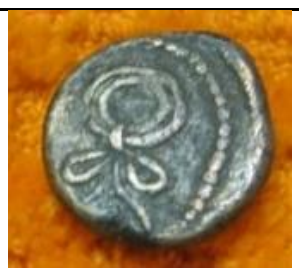 | Ibid | ? | 3/042 | 1.9 | copper | Coin of Phraates I I9 |
| 12 | 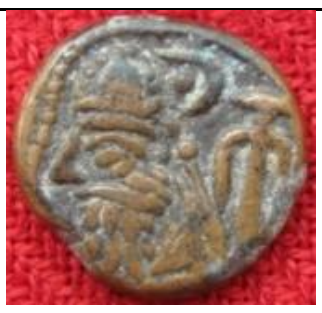 | 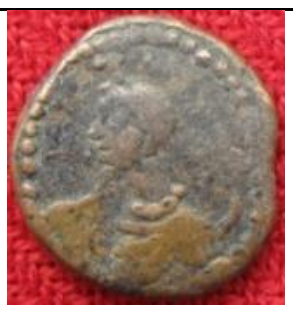 | Ibid | ? | 2/993 | 1.9 | copper | Coin of Orodes II F 6 |
| 13 | 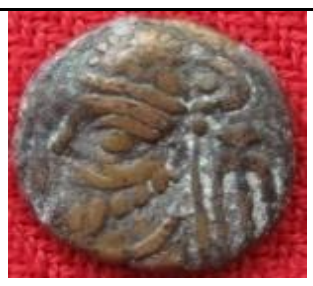 | 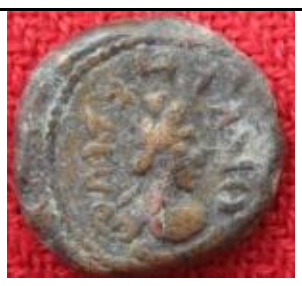 | Ibid | ؟ | 3/295 | 2 | copper | Coin of Orodes II E 5 |
| 14 | 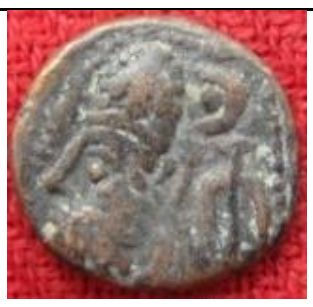 | 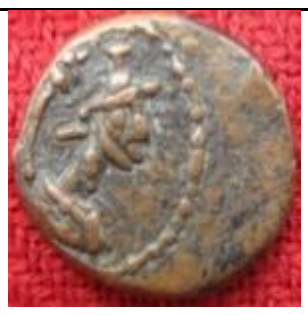 | Ibid | ? | 2/306 | 2.1 | copper | Coin of Orodes II B 2 |
| 15 | 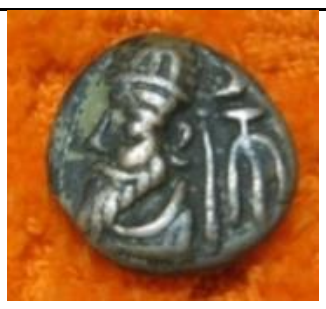 | 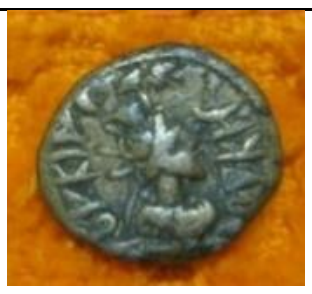 | Ibid | ? | 3/546 | 1.9 | copper | Coin of Orodes II M13 |
| 16 | 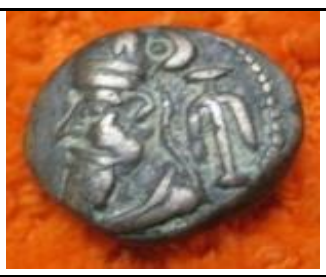 | 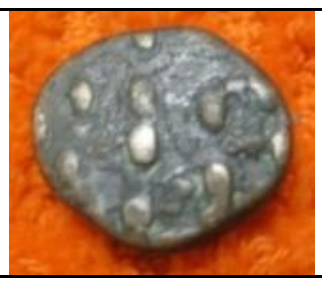 | | | | | | Coin of Orodes II |
| | | | Ibid | ? | 3/897 | 1.8 | copper | N14 |
| 17 | 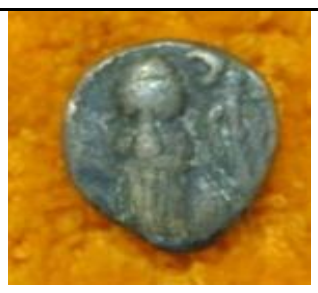 | 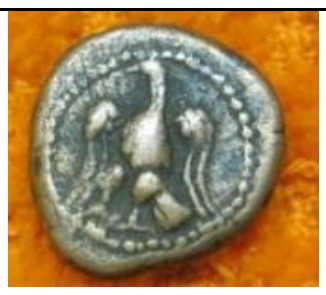 | reference | mint | Weight (gr) | Diameter (cm) | material | Coin of Phraates II G 7 |
| | | | Ibid | ? | 2/989 | 1.8 | copper | |
| 18 | 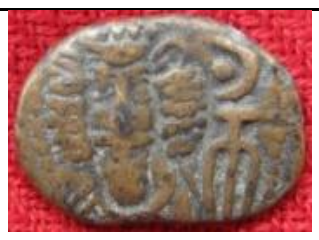 | 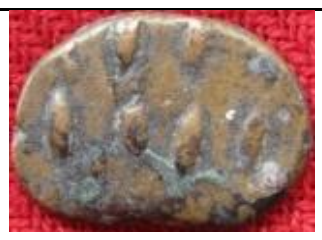 | reference | mint | Weight (gr) | Diameter (cm) | material | Coin of Kamnaskires-Orodes I D 4 |
| | | | Ibid | ؟ | 3/025 | 2.5 | copper | |
| 19 | 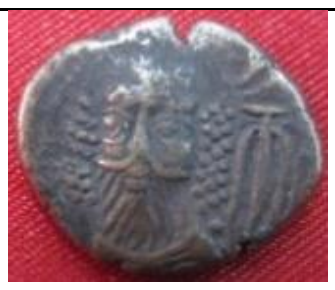 | 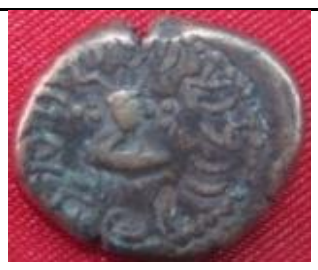 | reference | mint | Weight (gr) | Diameter (cm) | material | Coin of Kanaskires – Orodes I Q |
| | | | Ibid | ? | 3/290 | 2 | copper | |
| 20 | 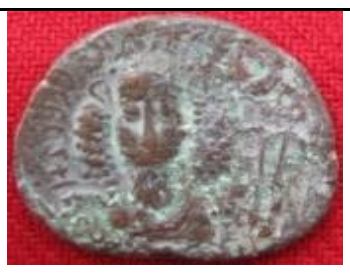 | 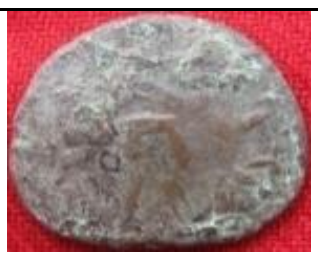 | reference | mint | Weight (gr) | Diameter (cm) | material | Coin of Kamnaskires-Orodes IW |
| | | | Ibid | ? | 13/801 | 3.2 | Bilon | |
| 21 | 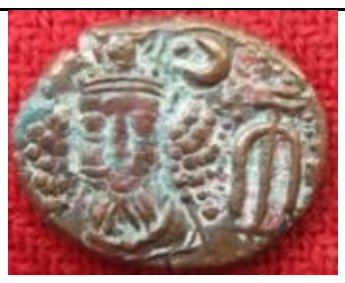 | 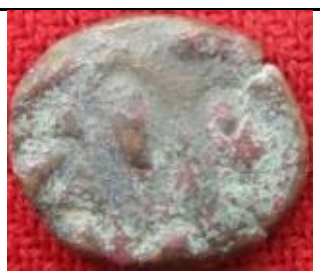 | reference | mint | Weight (gr) | Diameter (cm) | material | Coin of Kamnaskires Orodes II P 15 |
| | | | Ibid | ? | 4/096 | 2.3 | copper | |
| 22 | 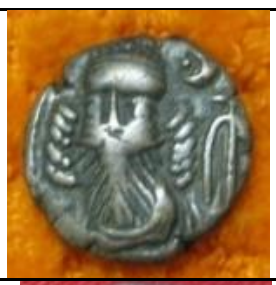 | 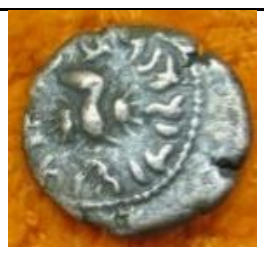 | reference | mint | weight (gr) | Diameter (cm) | material | Coin of Kamnaskires-Orodes II H 8 |
| | | | Ibid | ? | 2/713 | 1.9 | Bilon | |
| 23 | 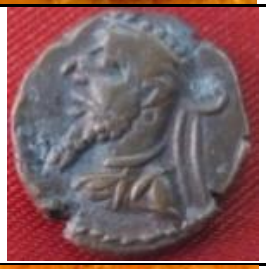 | 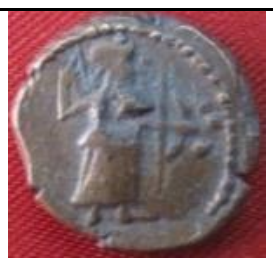 | reference | mint | Weight (gr) | diameter (cm) | material | Coin of Kamnaskires-Orodes II H |
| | | | Ibid | ? | 2/51 | 1.9 | copper | |
| 24 | 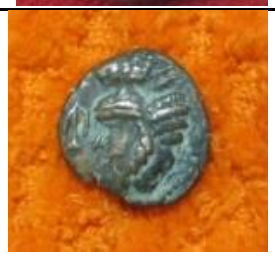 | 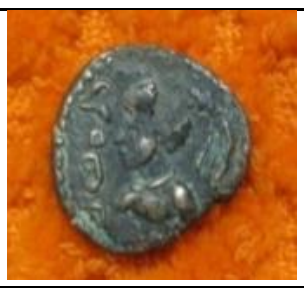 | reference | mint | Weight (gr) | Diameter (cm) | material | Coin of Orodes V and Olfan L |
| | | | Ibid | ؟ | 2/809 | 1.7 | copper | |
| 25 | 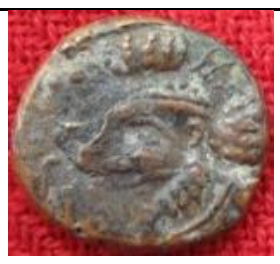 | 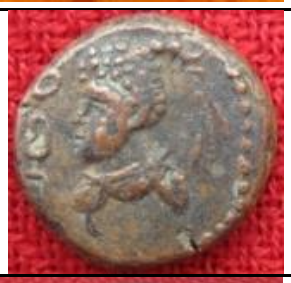 | refeernce | mint | weight (gr) | Diameter (cm) | material | Coin of Orodes V and Olfan K 11 |
| | | | Ibid | ? | 3/233 | 2.2 | copper | |
| 26 | 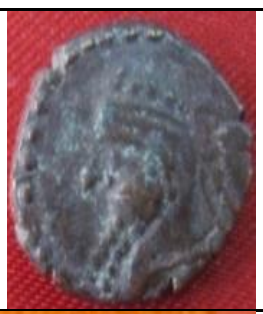 | 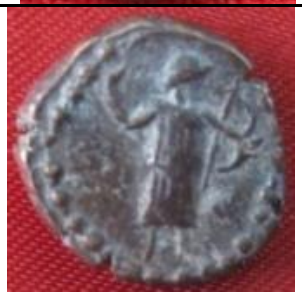 | reference | mint | Weight (gr) | Diameter (cm) | material | Coin of Orodes VII E |
| | | | Ibid | ? | 2/16 | 1.9 | copper | |
| 27 | 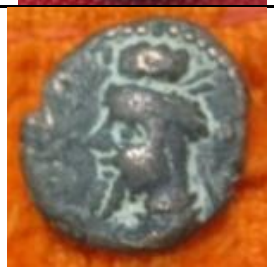 | 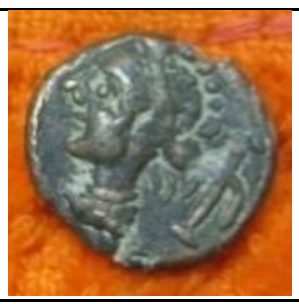 | reference | mint | weight (gr) | Diameter (cm) | material | Coin of Orodes VII J10 |
| | | | Ibid | ? | 2/727 | 1.9 | copper | |
| 28 | 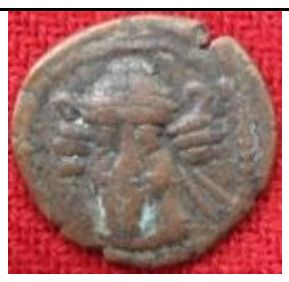 | 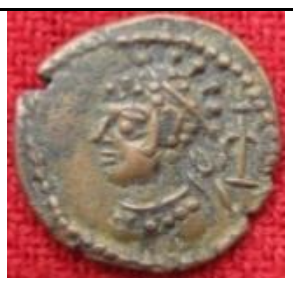 | reference | mint | Weight (gr) | Diameter (cm) | material | Coin of the third unknown king/Orodes VIII J |
| | | | Ibid | ? | 2/588 | 2 | copper | |
| 29 | 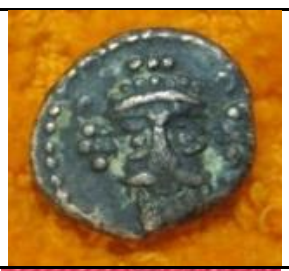 | 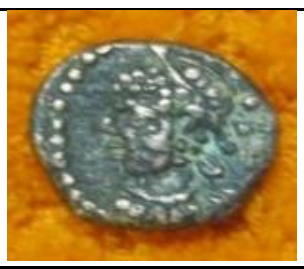 | reference | mint | Weight (gr) | Diameter (cm) | material | Coin of the third unknown king/ Orodes VII |
| | | | Ibid | ? | 2/739 | 2 | copper | |
| 30 | 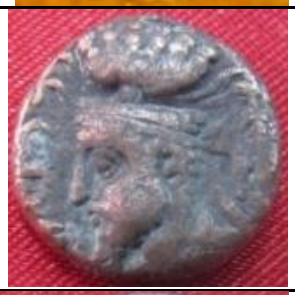 | 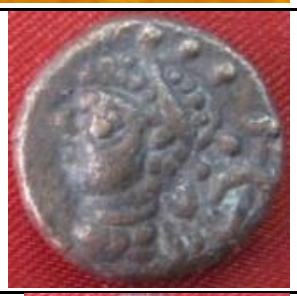 | reference | mint | Weight (gr) | Diameter (cm) | material | Coin of the third unknown kin/ Orodes VII gN |
| | | | Ibid | ? | 3/319 | 2.1 | copper | |
| 31 | 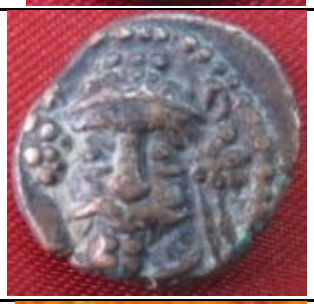 | 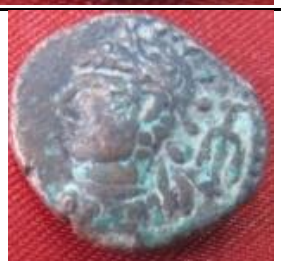 | reference | mint | Weight (gr) | diameter (cm) | material | Coin of the third unknown / Orodes VII ،B |
| | | | Ibid | ? | 2/93 | 1.9 | copper | |
| 32 | 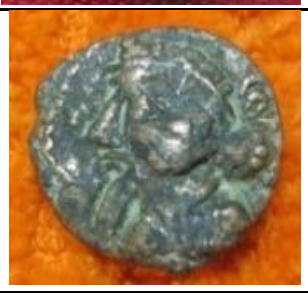 | 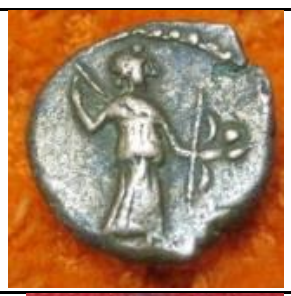 | reference | mint | weight (gr) | Diameter (cm) | material | Coin of the fifth unknown king E L12 |
| | | | Ibid | ? | 2/653 | 1.7 | copper | |
| 33 | 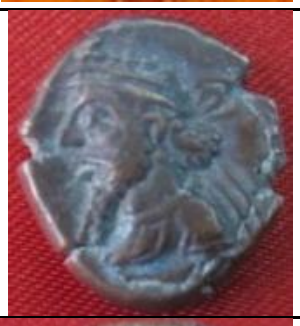 | 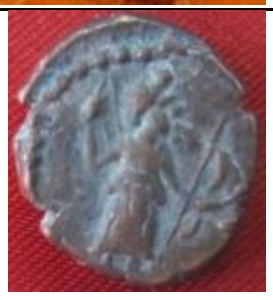 | مreference | mint | Weight (gr) | Diameter (cm) | material | Coin of the sixth unknown king C |
| | | | Ibid | ? | 2/784 | 2 | copper | |
| 34 | 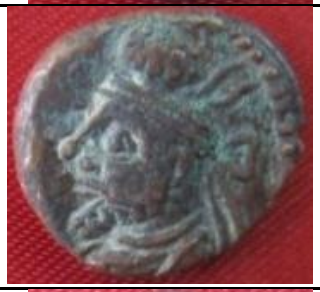 | 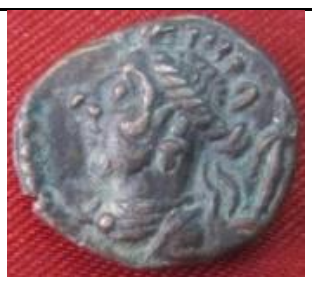 | referen ce | mint | Weight (gr) | Diameter (cm) | material | Coin of the sixth unknown king D |
| | | | Ibid | ? | 2/528 | 1.7 | copper | |
| 35 | 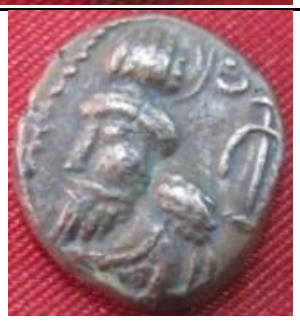 | 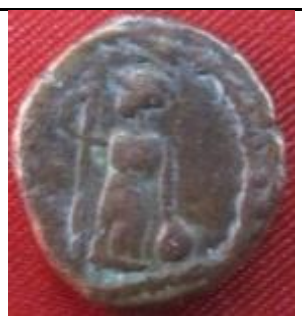 | referenc e | min t | Weight (gr) | Diamete r (cm) | material | Coin of the seventh P(G) unknown king |
| | | | Ibid | ? | 2/806 | 1.9 | copper | |

Table 2. The analyzed coins of Elymais and their characteristics

| Coin | Number of Coin | Timescope of Government | Elymais Kings |
|---|---|---|---|
| 1 | 1 | 82/81- 72/71 B.C | Kamnaskires Iv and oueen Anzaze |
| 2 | 1 | 50- 48 B.C | Kamnaskires VIII |
| 3 | 1 | 44-28 B.C | Kamnaskires XI |
| 4-6 | 3 | The late first century BC, Before Christ years | Kamnaskires XIII |
| 7-9 | 3 | The early first century AD, years before 30 AD | Other kings of Kamnaskires dynasty |
| 10 and 11 | 2 | Ca. 35- 50 AD | Phraates I |
| 12- 16 | 5 | Ca. 50- 70 AD | Orodes II |
| 17 | 1 | Ca. 70- 85 AD | Phraates II |
| 18- 20 | 3 | Ca. 100- 110 AD | Kamnaskires-Orodes I |
| 21- 23 | 3 | Ca. 110- 120 AD | Kamnaskires-Orodes II |
| 24 and 25 | 2 | Ca. 140- 150 AD | Orodes V and Olfan |
| 26- 27 | 2 | Ca. 155- 160 AD | Orodes VII |
| 28- 31 | 4 | The Secon half of first century AD | The third unknown king/ Orodes VIII |
| 32 | 1 | The Secon half of first century AD | The fifth unknown king (E) |
| 33 and 34 | 2 | The Secon half of first century AD/ Early third century AD | The sixth unknown king (F) |
| 35 | 1 | The Secon half of first century AD/ Early third century AD | The seventh unknown king (G) |

Table. 3: The Analyzed coins by PIXE technique (Authors)

| No | Sample | SI | CL | Ca | Ti | Fe | Ni | Cu | Zn | Ag | Sn | Pb | K | p | Al | others | King |
|---|---|---|---|---|---|---|---|---|---|---|---|---|---|---|---|---|---|
| 1 | A | 0.24 | 1.51 | 0.68 | . | 0.05 | . | 4.23 | . | 91.56 | . | 1.05 | . | . | . | Au:0.59 | Kamnaskires IV and Anzaze |
| 2 | I | . | 1.43 | 0.65 | . | . | 0.13 | 24.47 | . | 71.92 | . | 1.4 | . | . | . | . | Kamnaskires VII |
| 3 | O | 5.98 | 0.65 | 2.36 | 0.18 | 0.76 | 0.24 | 47.74 | . | 41.28 | . | 2.88 | . | . | 0.93 | . | Kamnaskires XI |
| 4 | V | . | 0.66 | 0.53 | . | 0.13 | 0.42 | 71.41 | . | 22.33 | 2.89 | 1.63 | . | . | . | . | Kamnaskires XIII |
| 5 | T | 0.25 | 0.68 | 0.29 | . | 0.06 | 0.49 | 81.66 | . | 11.1 | 3.47 | 2 | . | . | . | . | Kamnaskires XIII |
| 6 | G | 0.29 | 0.66 | 0.69 | . | 0.08 | 0.17 | 54.66 | . | 39.74 | . | 3.71 | . | . | . | . | Kamnaskires XIII |
| 7 | F | 0.68 | 0.31 | 0.61 | . | 0.13 | 0.46 | 90.82 | 0.56 | 0.93 | 2.71 | 2.79 | . | . | . | . | late Kamnaskires successors |
| 8 | S | 0.32 | 4.67 | 0.57 | . | 0.11 | 0.34 | 62.6 | . | 7.97 | 11.43 | 11.6 | . | 0.39 | . | . | lateKamnaskires Successors |
| 9 | R | . | 0.45 | . | . | 0.05 | 0.44 | 81.68 | . | . | 5.81 | 11.57 | . | . | . | . | lateKamnaskires Successors |
| 10 | C 3 | 3.51 | 1.44 | 1.08 | . | 0.16 | 0.38 | 62.69 | . | . | 1.69 | 29.05 | . | . | . | . | Phraates I |
| 11 | I 9 | 3.23 | 0.75 | 0.45 | 0.07 | 0.24 | 0.48 | 75.71 | . | . | 1.27 | 17.8 | . | . | . | . | Phraates I |
| 12 | F 6 | 4.15 | 1.61 | 1.45 | . | 0.28 | 0.37 | 57.05 | 0.23 | 3 | . | 31.62 | 0.47 | . | . | . | Orodes II |
| 13 | E 5 | 4.29 | 1.63 | 1.01 | . | 0.18 | 0.36 | 64.46 | 0.46 | . | 6.73 | 21.09 | . | . | . | . | Orodes II |
| 14 | B 2 | 4.16 | 0.91 | 2.05 | 0.05 | 0.28 | 0.48 | 79.31 | . | . | . | 12.48 | 0.05 | . | . | . | Orodes II |
| 15 | M13 | 4.47 | 1 | 0.46 | 2.91 | 0.06 | 0.41 | 0.43 | 73.16 | . | 2.97 | 14.13 | 0.46 | | . | . | Orodes II |
| 16 | N 14 | 6.77 | 3.14 | 2.65 | 0.07 | 0.56 | 0.31 | 54.71 | . | . | 10.23 | 21.56 | . | . | . | . | Orodes II |
| 17 | G 7 | 5.79 | 1.01 | 1.65 | . | 0.37 | 0.45 | 65.75 | . | . | 1.81 | 22.73 | 0.44 | . | . | . | Phraates II |
| 18 | D 4 | 5.77 | 0.98 | 1.88 | . | 0.51 | 0.37 | 73.88 | . | 1.3 | 7.82 | 7.03 | . | . | . | . | Kamnaskires - Orodes I |
| 19 | Q | 0.65 | 0.54 | 0.67 | . | 0.07 | 0.39 | 68.97 | 0.31 | 1.63 | 6.98 | 19.79 | . | . | . | . | Kamnaskires - Orodes I |
| 20 | W | 0.38 | 9.54 | 0.44 | . | 0.09 | 0.39 | 68.34 | . | 6.45 | 2.57 | 10.83 | | 0.36 | . | S: 0.63 | Kamnaskires - Orodes I |
| 21 | P 15 | 1.51 | 17.4 | 0.54 | . | 0.12 | 0.17 | 40.1 | . | . | 4.46 | 35.7 | . | . | . | . | Kamnaskires - Orodes II |
| 22 | H 8 | 3.65 | 0.94 | 0.65 | . | 0.12 | 0.36 | 69.3 | . | 4.33 | 5.75 | 14.9 | . | . | . | . | Kamnaskires - Orodes II |
| 23 | H | 0.42 | 1.19 | 0.3 | . | 0.81 | 0.33 | 63.31 | . | . | 2.72 | 30.92 | . | . | . | . | Kamnaskires - Orodes II |
| 24 | L | 0.24 | 0.47 | 0.26 | . | . | 0.39 | 77.27 | . | . | 3.58 | 17.79 | . | . | . | . | OrodesV&UlPhan |
| 25 | K 11 | 3.21 | 1.19 | 1.45 | . | 0.17 | 0.31 | 64.72 | . | . | 10.16 | 18.79 | . | . | . | . | OrodesV&UlPhan |
| 26 | E | 0.37 | 0.73 | 0.89 | . | 0.08 | 0.36 | 56.99 | . | . | 5.04 | 35.54 | . | . | . | . | Orodes VII |
| 27 | J 10 | 2.1 | 6.32 | 0.51 | . | 0.25 | 0.23 | 38.77 | . | . | 1.3 | 50.52 | . | . | . | . | Orodes VII |
| 28 | J | 1.06 | 0.56 | 2.07 | 0.1 | 0.26 | 0.68 | 74.61 | . | . | 2.41 | 18.25 | . | . | . | . | Third unkhown King (C) |
| 29 | A 1 | 4.66 | 1.53 | 0.95 | 0.08 | 0.44 | 0.31 | 55.97 | . | . | . | 35.59 | 0.37 | . | . | . | Third unkhown King (C) |
| 30 | N | 0.48 | 0.45 | 0.5 | . | . | 0.35 | 52.99 | . | . | . | 45.23 | . | . | . | . | Third unkhown King (C) |
| 31 | B | 1.54 | 0.35 | 1.93 | .. | 0.28 | 0.46 | 71.2 | 1.04 | . | . | 23.2 | . | . | . | . | Third unkhown King (C) |
| 32 | L 12 | 5.53 | 2.61 | 1.08 | . | 0.15 | 0.32 | 40.19 | . | . | . | 49.84 | 0.28 | . | . | . | Fifth unkhown King (E) |
| 33 | C | 0.38 | 1.83 | 0.47 | . | . | 0.46 | 77.9 | . | . | . | 18.96 | . | . | . | . | sixth unkhown King (F) |
| 34 | D | 0.78 | 0.62 | 0.52 | . | 0.13 | 0.41 | 70.83 | 0.5 | . | 7.61 | 18.6 | . | . | . | . | sixth unkhown King (F) |
| ۳۵ | P | 0.25 | 2.76 | 0.37 | . | . | 0.33 | 66.86 | . | . | 2.77 | 26.43 | 0.23 | . | . | . | Seventh unkhown King (G) |

Table.4: Available elements concentration in the Elymais coins by PIXE technique (Authors)

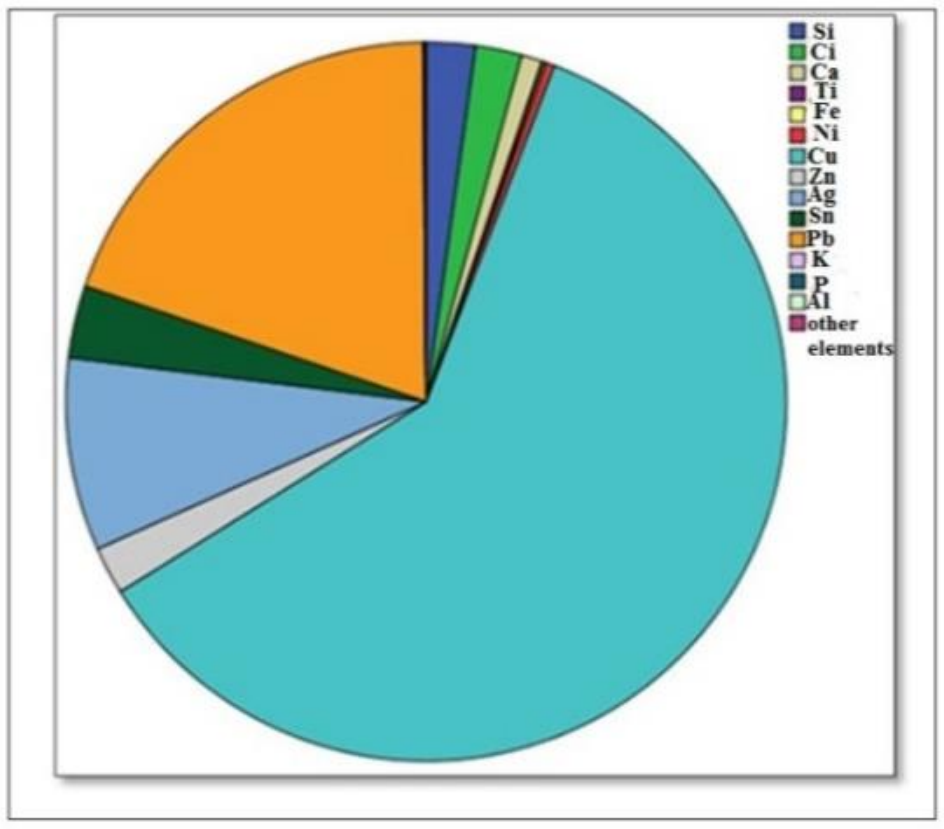

Figure. 1: The percentage of the used elements in the analyzed coins (Authors)

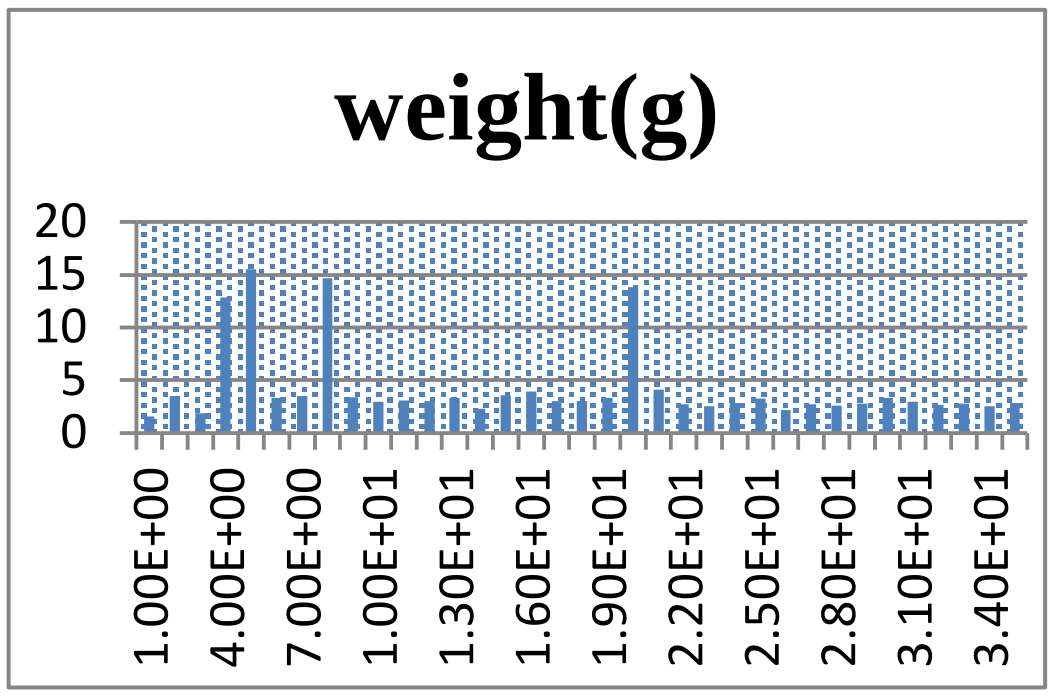

Figure.2: The weight of the analyzed coins of Elymais

Before the experiment, the analyzed coins were classified into two groups of silver and Bilon, but after analysis, coins were classified into three groups: 1. silver coins: coins which their silver was more than 50 percent, 2. Bilon coins: coins which their silver was between 3 and 49/9 percent, and 3. copper coins: coins which their silver was less than 3 percent (Hajvalaiei, personal correspondence). The increase of the lead more than 2 percent has been done to reduce the melting point of copper and it is mostly seen in the copper coins. Iron oxide and copper oxide represents that silver coins are counterfeit. Chlorine and calcium compounds represent that coins have long been underground or in the mintage process for preventing sticking of molten material into the mold have been pouring salt on the coins (Khademi Nadooshan & Moosvai, 2006).

The amount of tin (Sn) and zinc (Zn) in the silver coins was probably because of the shortage of bronze, brass and, copper (Cu) in the time of remelting. The copper element causes the hardening of coins. During the extraction of silver, copper remains less than 1 percent; The amount of copper more than 1 percent represents that coins have been minted in the various mines. The high percentage of iron (Fe) indicates that the added copper to silver has not been extracted well and iron is as the impurity. Gold (Au) is the associated element in the silver coins, but the lack of the element of gold in the silver coins cannot be indicated that the coin is fake, because in ancient times mines were used that did not have gold element. The existence of small amounts of lead (Pb) in the silver coins represents a great technique of extraction in that period. Gold, silver, copper and, lead are the heavy or original metals and titanium (Ti), iron, manganese (Mn) are not part of the heavy metals and are seen the slag or impurity on the surface of the coin (Hajvalaiei, 2009). In the classification of coins based on the place of coinage and identification of their mines, the combination containing copper and bismuth (Bi) can be used; especially when the archaeological features are ambiguous. Based on the elements of gold and iridium (Ir) can be identified as mines of coinage (Meyers, 1973) (Figures 3-7).

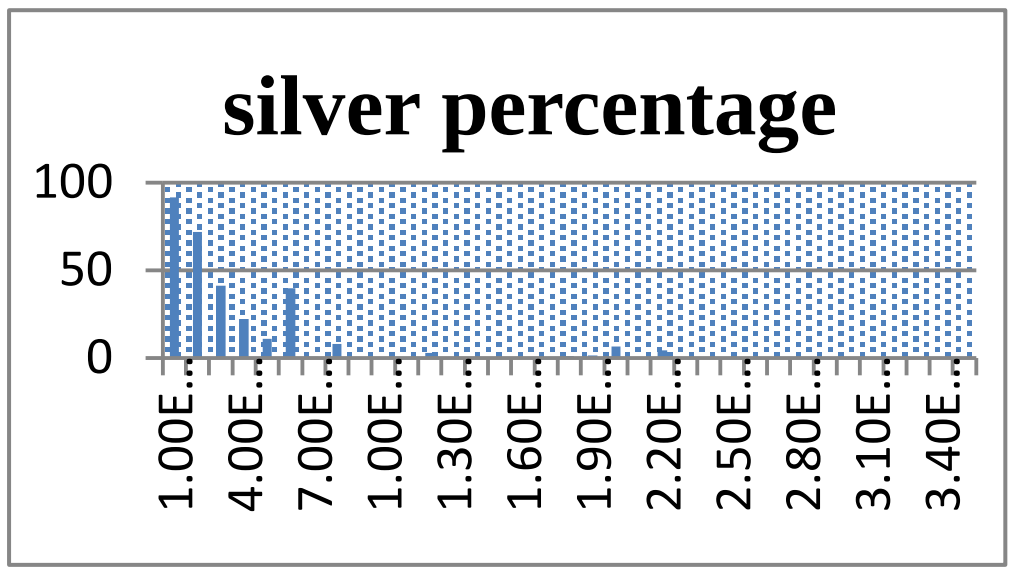

Figure. 3: Silver percentage of the analyzed coins

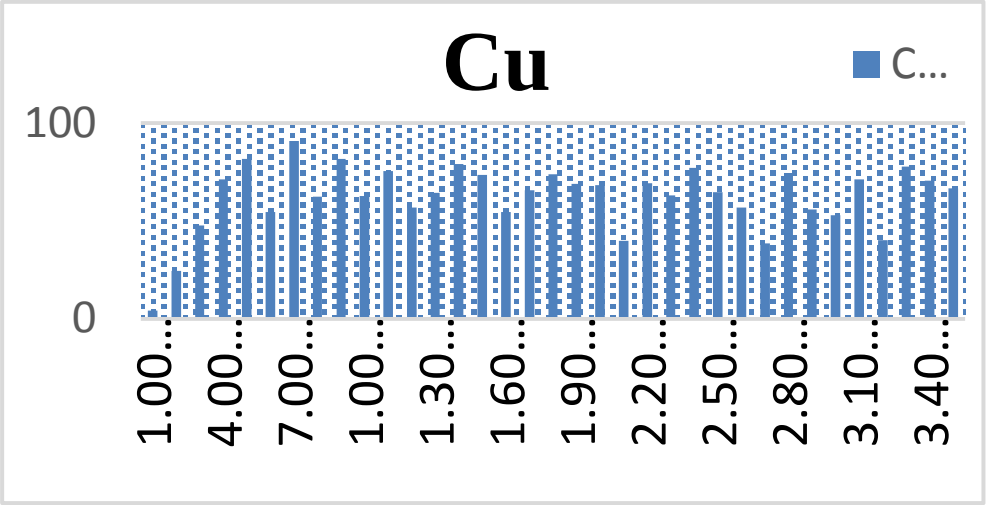

Figure. 4: Copper percentage of the analyzed coins

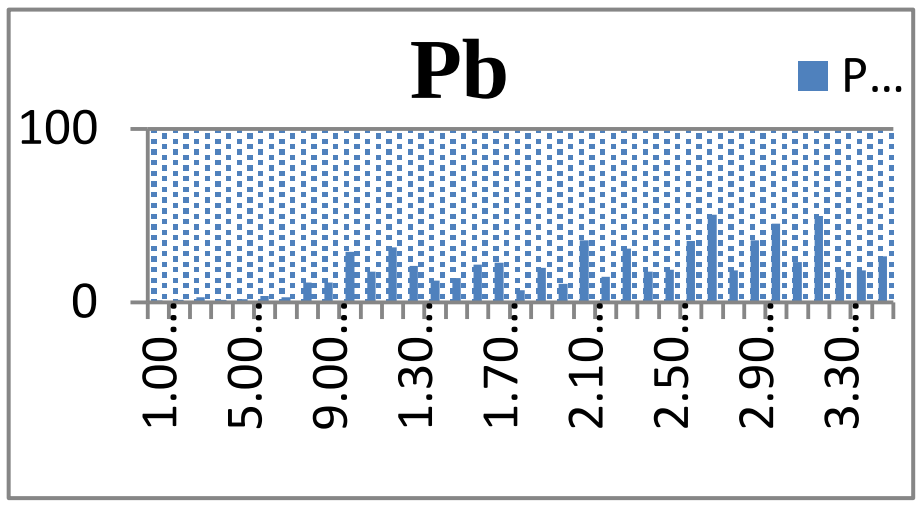

Figure. 5: Lead percentage of analyzed coin

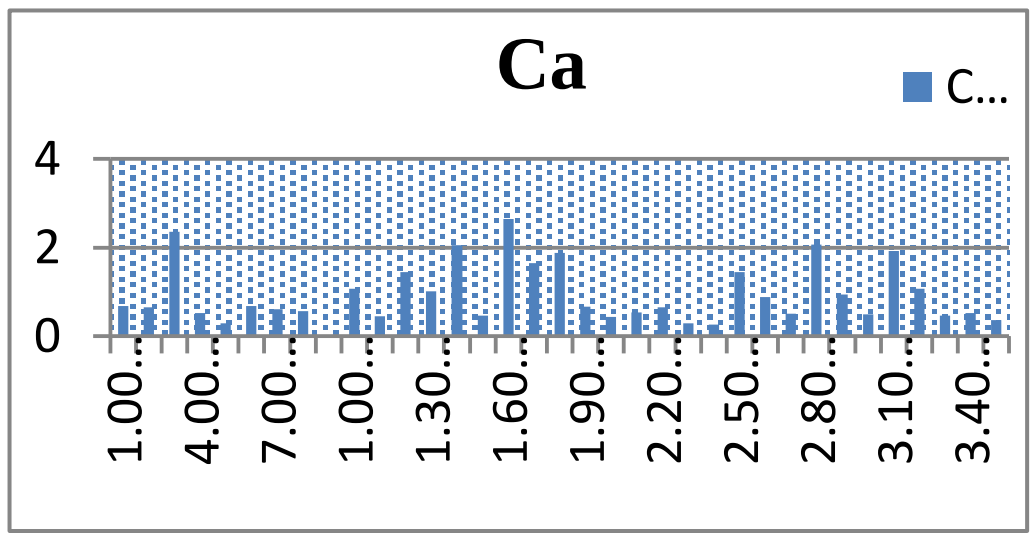

Figure. 6: Calcium percentage in the analyzed coins

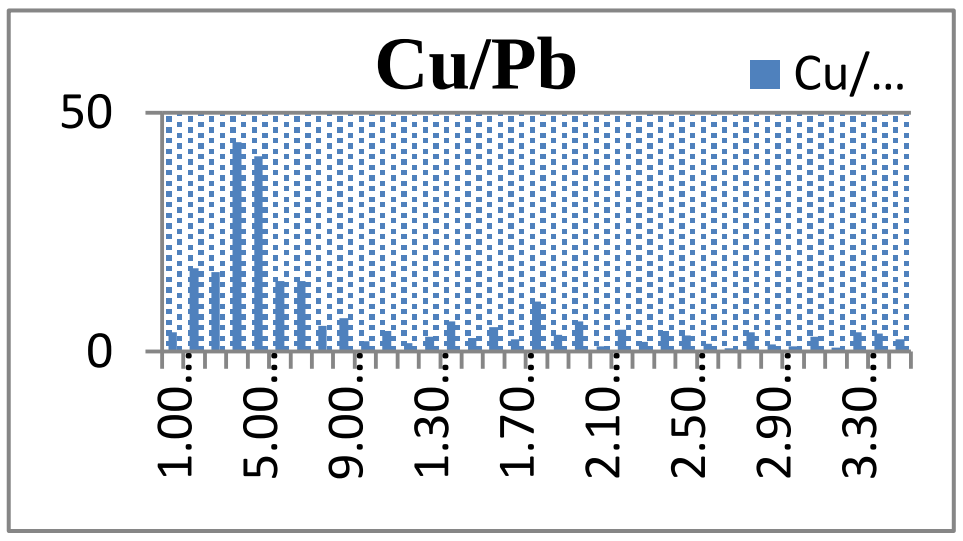

Figure. 7: The amount of copper to lead in the analyzed coins

Table 5 examines the amount of elements in these coins and the results are listed in terms of percentage. As shown, the elements of copper and lead are used in full (100 percent) in all 35 analyzed coins in this table, and in contrast, metals such as silver are used in 13 coins (37.1 percent). The amount of lead and silver in the coins is noteworthy. In this way, these two metals have been used in 100 percent of the coins to be analyzed.

| | Cases | | | | | |
|---|---|---|---|---|---|---|
| | Included | | Excluded | | Total | |
| | N | Percent | N | Percent | N | Percent |
| Silicon | 32 | 91.4% | 3 | 8.6% | 35 | 100.0% |
| Chlorine | 35 | 100.0% | 0 | 0.0% | 35 | 100.0% |
| titanium | 7 | 20.0% | 28 | 80.0% | 35 | 100.0% |
| Iron | 20 | 57.1% | 15 | 42.9% | 35 | 100.0% |
| Calcium | 34 | 97.1% | 1 | 2.9% | 35 | 100.0% |
| Nickle | 34 | 97.1% | 1 | 2.9% | 35 | 100.0% |
| Copper | 35 | 100.0% | 0 | 0.0% | 35 | 100.0% |
| Zink | 4 | 11.4% | 31 | 88.6% | 35 | 100.0% |
| Silver | 13 | 37.1% | 22 | 62.9% | 35 | 100.0% |
| Tin | 24 | 68.6% | 11 | 31.4% | 35 | 100.0% |
| Lead | 35 | 100.0% | 0 | 0.0% | 35 | 100.0% |
| Potassium | 7 | 20.0% | 28 | 80.0% | 35 | 100.0% |
| phosphorus | 2 | 5.7% | 33 | 94.3% | 35 | 100.0% |
| Aluminlum | 1 | 2.9% | 34 | 97.1% | 35 | 100.0% |
| Other element | 2 | 5.7% | 33 | 94.3% | 35 | 100.0% |

Table. 5: Comparison of the elements used in coinage based on their percent (authors)

In table 6, the high percentage of the use of these elements in the sections of mean and median is seen correctly, and in the standard deviation section, which examines the available fluctuations in the use of these elements, the most fluctuations in the use of copper (20.10027), zinc (36.29327), silver (29.59663) and lead (13.64781) metals are seen. This indicates the amount of differences in the use of these elements in the various periods of the Elymaean government. But in the case of other elements in these coins, which are not the main elements, these fluctuations are very small.

| | | Si | Cl | Ca | Ti | Fe | Ni | Cu | Zn | Ag | Sn | Pb | K | P | Al | other Element |
|---|---|---|---|---|---|---|---|---|---|---|---|---|---|---|---|---|
| N | Valid | 32 | 35 | 34 | 7 | 20 | 34 | 35 | 4 | 13 | 24 | 35 | 7 | 2 | 1 | 2 |
| | Missing | 3 | 0 | 1 | 28 | 15 | 1 | 0 | 31 | 22 | 11 | 0 | 28 | 33 | 34 | 33 |
| Mean | | 2.4097 | 2.0720 | .9606 | .4943 | .2145 | .3691 | 60.3309 | 18.7225 | 23.3492 | 4.7575 | 19.6286 | .3286 | .3750 | .9300 | .6100 |
| Median | | 1.5250 | 1.0000 | .6600 | .0800 | .1300 | .3700 | 64.7200 | .7500 | 7.9700 | 3.5250 | 18.6000 | .3700 | .3750 | .9300 | .6100 |
| Std. Deviation | | 2.17853 | 3.24988 | .66128 | 1.06607 | .19632 | .10581 | 20.10027 | 36.29327 | 29.59663 | 2.99852 | 13.64781 | .15334 | .02121 | | .02828 |

Table. 6: The percentage of elements in the analyzed coins (authors)

As seen in Table 6 and Figure 8, the ratio of the mean and the distribution of these two elements in the analyzed coins is very high, and this indicates the economic weakness during this period. Another indicator that is very important in the study of coinage technology and therefore the economic status of the study is the ratio of silver to lead, the lower ration indicates the use of the better technology to separate lead from the silver mine. As seen in Table 4, in the standard deviation section, as well as in figure 9, this distribution is very large and indicates the weakness of lead extraction technology from the silver mines. This distribution and inequality have been a sign of great fluctuations during the reign of the Elymaean kings, which had a profound effect on the coins of this period.

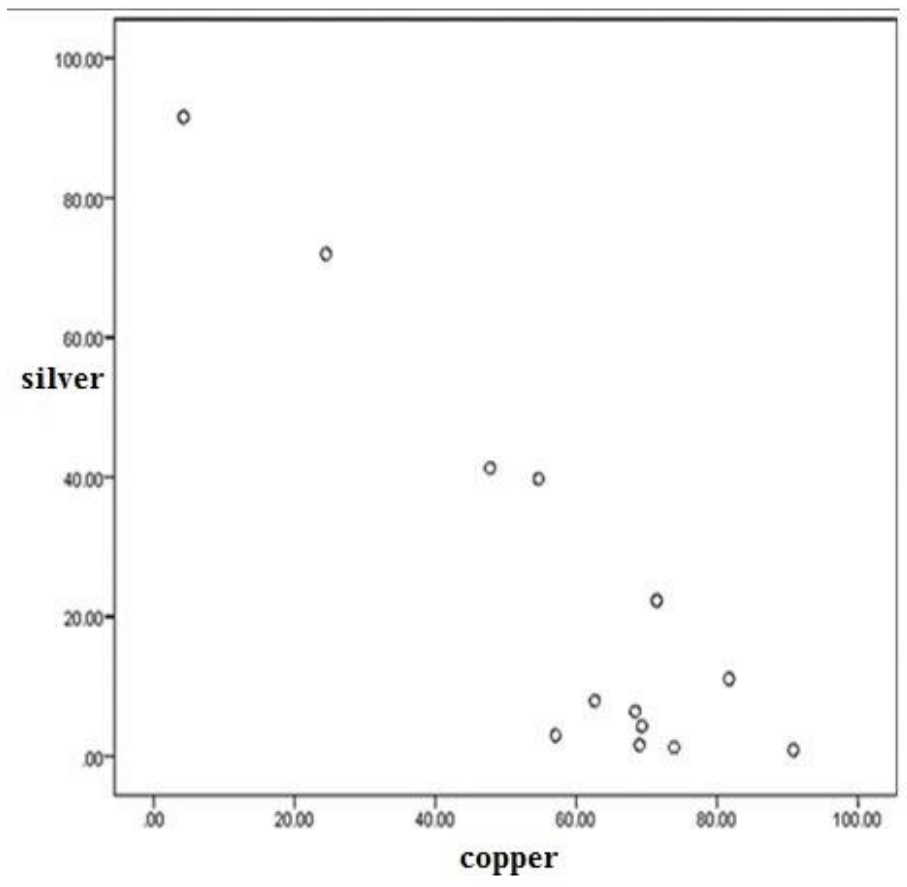

Figure. 8: The amount of silver to copper in the analyzed coins

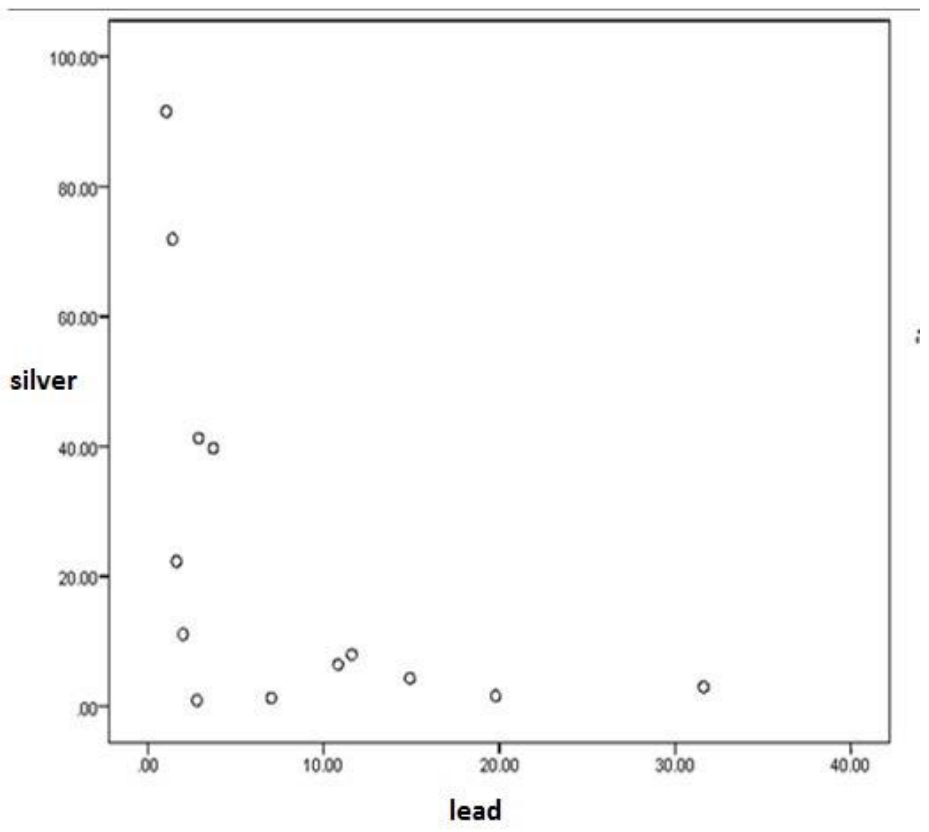

Figure. 9: The amount of silver to lead in the analyzed coins

## Conclusion:

The analyzed coins of Elymais can be studied in three-time scope. The first period is related to the Kamnaskires dynasty that it includes coins from 80 BC to 25 AD. From the second and third periods coins of the various kings analyzed. It almost includes the maximum time scopeof these two periods. The analyzed coins of Elymaean are made of silver (2), copper (26), and Bilon (7). Of the percentage of available elements of the analyzed coins is used to understand the economic situation, the number of mints and mines. The percentage of silver in the coins of the first kings is high, but there was no silver during the second and third periods or it was very rare (Figure. 3). This subject can be indicated a better economic situation during the first kings of Elymais than the second and third periods. Two coins of 35 analyzed coins of Elymais are silver (Coins of 1 and 2), the coin belonging to Kamnaskires IV, and its silver percentage is high. Although one or two coins are not enough for analysis, it can be said increasing the percentage of silver and the small percentage of the coins impurity indicate that the economic situation of the reign of this king was suitable and reducing the amount of silver and increasing of the impurity in the reign of Kamnaskires VIII indicate that the economic situation was bad. In addition to copper, the silver percentage in the coins of Kamnaskires XI and Kamnaskires XIII are high. Bilon coins have had more value than copper coins or because of the precipitance in the coinage and metallurgy weakness in this period, they have not good quality. Also, there was a small percentage of silver in the copper coins during the other kings of the Kamnaskires dynasty. About this subject of whether this style of coinage has been used during the first kings of Elymais or not, we can not say certainly, but in the copper coins, we can see this style. During the Kamnaskires- Orodes I and Kamnaskires- Orodes II, we can see the percentage of silver in the copper coins again. Since the percentage of copper to lead in these coins are different, this can be confirmed by different mints and may be proposed the popularity of this style of the coinage (Bilon) for Elymaean. After the reign of the first kings of Elymais, this style cannot be seen during the reign of the Parthian successors. In the time of Kamnaskires I who Susa and Elymais perfectly linked together, Bilon is seen once again. It can be said this style of coinage was a native and local style for the mountain dwellers of Elymais. So that, Elymaean in the plains of Khuzestan have diffused this style. This ethnic duality in the history of the area and even pre-history can be studied. Of course, we need the analysis of the more coins of Elymais to confirm or reject this hypothesis. Also, the subject of the superficial difference between this type of coins and copper coins requires an independent investigation.

The copper percentage in the copper coins of the first Elymais kings is more than the second and third periods, and amount of the copper in the coins of the second period has more stability (Figure. 4). The percentage of copper is increased in the coins of KamnaskiresVII. Also, the percentage of copper is increased in the Bilon coins of Kamnaskires XI, Kamnaskires XIII and, the other kings. The amount of copper in the copper coins of the first period is more than 80 percent and the percentage of impurity and lead is rare. In the second period of Elymais, the percentage of copper in the copper coins is between 50 and 80 percent. The amount of copper in the third period of Elymais has been between 35 and 80 percent. Thus, the percentage of copper in the coins has had a decreasing trend, that this subject can reveal economic changes in these three periods (more purity of the copper coins indicates the strong economy of that period).

The low lead percentage indicates the good metallurgy. Low melting temperature and resistance to the corrosion are good properties of this metal for coinage. In the silver and billion coins of the first period of Elymais, the percentage of lead is low, however, at the end of this period, the lead percentage increased. In the second period of Elymais, lead content has been between 10 and 30 percent, which in the third period arrived between 10 and 50 percent. This lead increase indicates changes in the three periods that it has been because of the diminution of available copper mines or the need for the rapid coinage and or remelting of coins (Figure. 5). Of the analyzed coins of this study, We can not speak about the exact location of their coinage; because the signs of mints are uncertain on them. Based on the coins analysis and with respect to the copper percentage to lead in the coins can be said the analyzed coins have been minted in six mints. The proportion of the percentage of the mentioned elements is considered with the tolerance of 1.

The percentage of copper to lead in the coins of the first period kings is more than the second and third periods. The amount of copper to lead has been in the three periods including the first period between 10 and 45 percent, the second period between 1 and 10 percent, and the third period between 1 and 5 percent. The coins of 7, 6, 3, 2 have been related to one mint. The coins of 34, 33, 31, 28, 25, 24, 22, 19, 16, 13, 11, 8, 1, the coins of 20, 14, 9, the coins of 35, 31 , 33, 30, 29, 26, 25, 23, 21, 19, 17, 15, 13, 12, 10, the coins of 32, 30, 29, 27, 26, 21, 12 and the coins of 5 and 4, each of them together have been minted in different mints. Some of the coins of the Elymais kings that their mints like together include: The coins of 8 and 11 belonging to the other lineage of Kamnaskires, the coins of 13 and 16 related to Orodes II, the coins of 24 and 25 attributed to Orodes V and Olfan, the coins of 12, 13 and 15 related to Orodes II, the coins of 21 and 23 related to Kamnaskires – Orodes II, the coins of 20, 30 and 31 belonging to the third unknown king and the coins of 33, 31 and 28 are related to the same Shah that this may be confirmed the coinage of this king in two mints (Table. 7).

Given the percentage of calcium in the coins of the first period Kings, can be divided the coins of the first period kings into two groups, the coins of the second-period kings into three and the coins of the third period Kings into two groups (figure. 6). Totally, based on the percentage of calcium, 35 coins can be divided into three categories: between 0 and 1, between 1 and 2 percent, and between 2 and 3 percent. According to this matter, can be stated that Elymaean have used three different mines to extract ore to the production of these coins. Unfortunately, we can not talk now about their name and place. To determine the mines, different mines should be sampled and analyzed separately to identify the used mines by comparative study. Based on the elemental analysis of these coins can be concluded that during the first-period kings of Elymais (Kamnaskires and his successors) toward the kings of the second period (Parthian dynasty) and the third period of Elymais (last kings) coinage had more important and the amount of coin impurity was lesser. This subject implies that the political and economic conditions of the first-period kings of Elymais have been better than the later periods of this government.

| Mint A | Mint B | Mint C | Mint D | Mint E | Mint F |
|---|---|---|---|---|---|
| Coins No.<br><br>7, 6, 3, 2 | Coins No.<br><br>33, 31, 28, 25, 24, 22, 19, 16, 13, 11, 8, 1 | Coins No.<br><br>20, 14, 9 | Coins No.<br><br>30, 31, 33, 29, 26. 25, 23, 21, 19, 17, 15, 13, 12, 10 | Coins No.<br><br>32, 30, 29, 27,26, 21, 12 | Coins No.<br><br>5, 4 |

Table. 7: The hypothetical location of 35 analyzed coins that were probably multiplied in a mint

.

**Acknowledgments:**

We whould like to thank the respectful authorities of Van de Graaff laboratory of Atomic Energy Organization of Iran (AEOI), Mr. Mohammad Lamei Rashti and Mrs. Parvin Ovliyaei who accepted the coins analysis of this research. Also, We would like to thank Mr. Mohammad Saffar who provided the collection of his coins for this research.